\documentclass{an}
\usepackage{times}
\usepackage{graphicx}
\usepackage{txfonts}
\newcommand{\ubv}{$UBV~$}
\newcommand{\ubvr}{$UBVR_{\rm C}~$}
\begin{document}
%

\Pagespan{1}{}


\Yearpublication{2007}%
\Yearsubmission{2007}%
\Month{03}%
\Volume{999}%
\Issue{88}%
\DOI{This.is/not.aDOI}%

\title{Recent photometry of symbiotic stars - XII\thanks{
       Tables 1--18 are also available in electronic form 
       at http://www.astro.sk/$\sim$astrskop/}}

\author{A. Skopal\inst{1}\fnmsep\thanks{Corresponding author:
        \email{skopal@ta3.sk}\newline}
\and  M. Va\v{n}ko\inst{1}
\and  T. Pribulla\inst{1}
\and  D. Chochol\inst{1}
\and  E. Semkov\inst{2}
\and  M. Wolf\inst{3}\thanks{Visiting Astronomer, South African 
                             Astronomical Observatory}
\and  A. Jones\inst{4}
}
\titlerunning{Recent photometry of symbiotic stars}
\authorrunning{A. Skopal et al.}
\institute{
     Astronomical Institute, Slovak Academy of Sciences,
     059\,60 Tatransk\'{a} Lomnica, Slovakia
\and
     Institute of Astronomy, Bulgarian Academy of Sciences,
     Tsarigradsko shose Blvd. 72, Sofia 1784, Bulgaria
\and
     Astronomical Institute, Charles University Prague, CZ-180 00
     Praha 8, \mbox{V Hole\v{s}ovi\v{c}k\'ach} 2, The Czech Republic
\and
     Carter Observatory, PO Box 2909, Wellington 1, New Zealand
}

\received{15 March 2007}
\accepted{}
\publonline{later}


\abstract{%
 We present new photometric observations of 15 symbiotic stars 
 covering their last orbital cycle(s) from 2003.9 to 2007.2. 
 We obtained our data by both classical photoelectric and CCD 
 photometry. Main results are:
EG\,And 
brightened by $\sim$\,0.3\,mag in $U$ from 2003. 
A $\sim$0.5\,mag deep primary minimum developed in 
the $U$ light curve (LC) at the end of 2006. 
Z\,And 
continues its recent activity that began during the 2000 
autumn. A new small outburst started in summer of 2004 with 
the peak $U$-magnitude of $\sim 9.2$. During the spring 
of 2006 the star entered a massive outburst. It reached 
its historical maximum at $U\sim 8.0$ in 2006 July. 
AE\,Ara 
erupted in 2006 February with $\Delta m_{\rm vis} \sim 1.2$\,mag. 
BF\,Cyg 
entered a new active stage in 2006 August. 
A brightness maximum ($U\sim 9.4$) was measured during 
2006 September. 
CH\,Cyg 
persists in a quiescent phase. During 2006 
June -- December a $\sim 2$\,mag decline in all colours was measured.  
CI\,Cyg 
started a new active phase during 2006 May -- June. 
After 31 years it erupted by about 2\,mag in $U$. 
TX\,CVn 
maintains a bright stage with $U \sim 10.5$ from 2003. 
AG\,Dra entered a new major outburst in 2006 June. It reached
its maximum at $U \sim 8.0$ in 2006 September. 
%
%
AR\,Pav 
persists at a low level of the activity. 
%
%
AG\,Peg's 
LC profile varies markedly during different orbital cycles.
%
AX\,Per 
continues its quiescent phase. 
}
\maketitle

\section{Introduction}

Symbiotic stars are long-period ($P_{\rm orb} \sim 1 - 3$ years) 
interacting binary systems consisting of a late-type 
giant and a hot compact star accreting from the giant's wind. 
This process generates a very hot 
   ($T_{\rm h} \approx 10^5$\,K)
and luminous 
   ($L_{\rm h} \approx 10^2 - 10^4\,L_{\sun}$)
source of radiation that ionizes a fraction of the neutral wind 
from the giant giving rise to nebular emission. 
As a result the observed spectrum of symbiotic stars composes 
from three basic components of radiation -- two stellar and one 
nebular. Throughout the optical their contributions rival 
each other, producing the composite spectrum, whose colour 
indices differ significantly from those of standard stars. 
In addition, they are different for individual objects and 
variable due to activity and/or the orbital phase
(cf. models SED in Skopal 2005). 
Therefore the LCs of symbiotic binaries bear a great deal 
of information about properties of the radiative sources 
in the system. Photometric monitoring is important 
to complement other multifrequency observations, mainly 
during outbursts, to improve our understanding of the 
observed phenomena (e.g. Sokoloski 2003). 

In this paper we present results of our long-term moni\-toring 
programme of photometric observations of selected symbiotic 
stars, originally launched by Hric \& Skopal (1989). It continues 
the work of Skopal et al. (2004, hereafter S+04) by collecting new 
data obtained during the period 2003 December to 2007 January. 
Their acquisition and reductions are introduced in Sect.~2. 
In Sect.~3 we note the most interesting features 
of the LCs that deserve further investigation. The results 
are presented in Tables~1--18 and Figs.~1--12. 

\section{Observations and reductions}

Observations made at the Skalnat\'{e} Pleso (hereafter SP 
in Tables), Star\'{a} Lesn\'{a} (G2 pavilion) and the Rozhen 
Observatory with the Schmidt telescope (R) were already 
described by S+04. Further details about the observation 
procedure were described by Hric et al. (1991). 
Other \ubv photoe\-lec\-tric observations were obtained with 
the modular photometer utilizing a Hamamatsu EA1516
photomultiplier on the 0.5-m telescope at the Sutherland site
of the South Africa Astronomical Observatory (SAAO) during
two weeks in Ap\-ril and May 2004 and September 2005.
The photoelectric measurements were done in the \ubv filters
of the Johnson's photometric system with a 20 second integration
time. Observations were reduced to the Cousins E-region 
standard system (Menzies et al. 1989) and corrected for 
differential extinction using the reduction program 
HEC~22 rel.~14 (Harmanec \& Horn 1998). 
Observations of Draco\,C-1 in the standard Johnson-Cousins system 
were made with the photometric AT-200 CCD camera 
(1024$\times$1024\,px, pixel size: 24$\times$24\,$\mu$m, scale: 
0.33 arcsec/px, field: 5.6$\times$5.6 arcmin) on the 2-m telescope 
at the Rozhen Observatory. 
Additional $BVR_{\rm C}I_{\rm C}$ CCD photometry was obtained 
with the 0.5-m telescope at the Star\'{a} Lesn\'{a} Observatory 
(G1 pavilion). The {\sf SBIG ST10 MXE} CCD camera with the chip 
2184$\times$1472 pixels was mounted at the Newtonian focus. 
The size of the pixel is 6.8\,$\mu$m and the scale 0.56$\arcsec$/pixel. 
All frames were dark-subtracted, flat-fielded and corrected for 
cosmic rays. Other details of the CCD photometric reduction were 
described by Parimucha \& Va\v{n}ko (2005). 

In addition, visual magnitude estimates of AE Ara, RW Hya and 
AR Pav were obtained by one of us (AJ) with a private 12".5 f/5 
reflector. A comparison between $V$ magnitudes and corresponding 
visual estimates suggests their uncertainties to be within about 
0.2\,mag for brighter objects (RW\,Hya, AR\,Pav) and 0.3\,mag 
for AE\,Ara during quiescence ($m_{\rm vis} \ga 13$\,mag). 

We measured our targets with respect to the same standard stars 
as in our previous papers (e.g. S+04) if not specified otherwise 
in Sect.~3. Results are summarized in Tables~1--18 and shown in 
Figs.~1--12. 
Each value represents the average of the observations during 
a night. The maximum internal uncertainty of these night-means 
is less than 0.05\,mag. We verified the absolute photometry by 
comparing our data with those obtained independently at other 
observatories. 
%

\section{Notes to measured objects}

%
\subsection{EG\,And}

EG\,And is a quiet symbiotic star -- no outburst of 
the Z\,And type has been recorded to date. 
Photometric measurements are listed in Table~1 and 
plotted in Fig.~1. From 2003 the $U$-LC indicates a brighter 
stage of EG\,And by about 0.3\,mag. According to the ephemeris 
of Skopal (1997), a deep primary minimum ($\Delta U \ga 0.5$\,mag) 
was observed during 2006 October to December. 
By a precise modeling the UV/optical continuum, Skopal (2005)
showed that the brightening in the $U$ passband can be caused 
by an increase in the nebular emission. This could be a result
of a transient increase in the mass-loss rate from the giant 
and thus also the accretion rate and the luminosity of the 
hot component. Consequently this process increases flux 
of ionizing photons that gives rise the larger amount of 
the nebular radiation. 
%
%

\subsection{Z\,And} 

Z\,And is a prototype symbiotic star. The star BD+47\,4192 
($V$ = 8.99, $B-V$ = 0.41, $U-B$ = 0.14, $V-R_{\rm C}$ = 0.10) 
was used as the comparison for both photoelectric and CCD 
observations (Tables~2 and 3, Fig.~2). They revealed two new 
eruptions. The first one started in 2004 July/August and 
peaked at $V \sim U \sim$\,9.2\,mag in the mid September with 
a following re-brightening in 2004 December. Then a slow 
decrease in the star's brightness was observed to the end 
of 2005. 
The second major eruption started during the spring of 2006 
and peaked in 2006 July, when the star's brightness reached 
its historical maximum ($U\sim 8.0$, $V\sim 8.5$) that has 
ever been recorded by the multicolour photometry. 
It is of interest to note that spectral features indicating 
ejection of highly collimated bipolar jets developed during 
the optical maximum (Skopal \& Pribulla 2006). Their evidence 
in the optical spectra was confirmed by 
Burmeister \& Leedj\"arv (2007) and Tomov et al. (2007). 
%
%

\subsection{AE\,Ara}

Our observations of AE\,Ara 
consists of 488 visual estimates made by one of us (AJ) carried out 
from 1987 to 2006.9 and photoelectric \ubv measurements made 
on 25/04/2005 at SAAO 
(JD\,2\,453\,122.596: $V$ = 12.402, $B-V$ = 0.607, $U-B$ = -1.044)
We used the standard star HD\,317858 
(CD-32\,12919; $V$ = 9.533, $B-V$ = 0.143, $U-B$ = -0.523). 
Results are shown in Fig.~3. There is a good agreement 
between the visual estimates and our photoelectric $V$ magnitude. 
The very negative $U-B$ index of AE\,Ara suggests 
a strong contribution from the nebula at the $U$ passband. 
During the 2000-05 period a wave-like variation developed in 
the LC. The time of the best defined minimum at 
JD~2\,453\,474$\pm$20 agrees (within the uncertainties) with 
that predicted by the ephemeris of Mikolajewska et al. (2003). 
Also the time of the previous minimum, we estimated to 
$\approx$JD~2\,452\,710, is close (within 0.07 of the orbital 
period) to the predicted one. 
This implies that this light variation was due to the orbital 
motion (see Mikolajewska et al. 2003 in detail). 
We note that a large scatter in our visual magnitudes did not 
allow us to estimate the position of the first minimum more 
accuratelly. 

During 2006 February AE\,Ara entered a new active phase. 
Our visual observations revealed a rapid increase in the star's 
brightness by about 1.2\,mag. It peaked at $m_{\rm vis} \sim 11.2$ 
during April and was gradually decreasing to $\sim$11.8 
in 2006 November before its season observational gap. 
%
%
\subsection{BF\,Cyg}

The resulting night-means of the BF\,Cyg 
brightness are in Table~4. Figure~4 shows its \ubv\ LCs covering 
the last 3 orbital cycles. 
The maximum between 2004.5 and 
2005.5 was complex in profile. First a 0.5\,mag increase with 
respect to values from previous cycles was observed during the 
second half of 2004. An additional brightening to $U\sim$10.5\,mag 
was observed in the spring of 2005. 
The following minimum at JD~2\,453\,705$\pm$12 was by $\sim$0.5\,mag 
brighter than those observed previously. 
In 2006 August the LC revealed an eruption with the peak 
$U$-magnitude of $\sim$9.4 during the following September. 
The active phase continues with a slow fading to our last 
observations at the end of 2006. 
We note that a similar profile of the LC was also observed
during the 1987-89 period, prior to the 1989 outburst 
(cf. Fig.~2 of Skopal et al. 1997). 
%

\subsection{CH\,Cyg}

Our new photometry of CH\,Cyg 
is listed in Table~5. Figure~5 shows LCs from the last 
1998-00 activity. From the beginning of 2000 CH\,Cyg persists 
in a quiescent phase at rather bright magnitudes ($V = 7\div8$, 
$B = 8.7\div9.4$ and $U \approx\,10$ or less). 
The LCs display a wave-like 750$\div$770-day periodic variation, 
more pronounced in $V$, whereas in $U$ the brightness only 
fluctuated around 10 from about 2003. This suggests that 
mainly a giant star in the system is responsible for such 
behaviour. This type of the LC profile developed 
during each previous post-outburst stage, in 1970 and 1987 
(see Fig.~1 of Eyres et al. 2002). 
During the 2006 June -- December period the LCs showed 
a 2\,mag decline in all colours 
($\Delta U \sim 1.8,~\Delta B \sim 2.3,~\Delta V \sim 2.5$\,mag). 
%

\subsection{CI\,Cyg}

Photometric measurements of CI\,Cyg 
are introduced in Tables~6 and 7 and depicted in Fig.~6. 
The wave-like variation along the orbital motion indicates 
a quiescent phase. Such the behaviour developed in 1985, about 
10 years after the last outburst in 1975 (cf. Dmitrienko 2000), 
and continued until the spring of 2006, when a new eruption 
was detected. A pre-outburst activity was indicated during 
the recent 2003-06 cycle when the $U$ star's brightness 
of the 2005-maximum was by about 0.4\,mag higher 
than that of the previous one (Fig.~6). Additional variations 
were recorded in the $V$ and (in part) $B$-band LCs. 
They are probably caused by the red giant whose light 
dominates these passbands during quiescence 
(see Fig.~10 in Skopal 2005).

During the 2006 May -- June period CI\,Cyg 
started its new active phase, when brightened by 
$\Delta U \sim 2$\,mag, 
$\Delta B \sim 1.2$\,mag and 
$\Delta V \sim 1$\,mag. 
%

\subsection{V1329\,Cyg}

Observations of V1329\,Cyg (HBV\,475) are given in Table~8. 
%

\subsection{TX\,CVn}

Table~9 and Fig.~7 summarize photometric measurements of 
TX\,CVn. 
From 2003 the system persists at a higher level of activity 
with $U \sim 10.5$. 
Sometimes during the brighter stages, the $U$-LC shows minima 
placed at the inferior conjunction of the cool giant 
(according to the ephemeris of Kenyon \& Garcia 1989 for 
a circular orbit; thick arrows in Fig.~8). However, in some 
cases no minima were detected in spite of a sufficient 
coverage of the corresponding part of the LC 
(crosses in Fig.~8). 
Another peculiarity concerns to the minima width. 
The minima are too broad than to be explained by the eclipse 
of simple stellar photospheres 
(e.g. $t_3 - t_2 \approx 0.2\,P_{\rm orb}$). 
This effect deserves further investigation. 
%

\subsection{AG\,Dra}

Our measurements of AG\,Dra 
are summarized in Tables~10 and 11 and plotted in Fig.~8. 
The stars "a" ($\alpha_{2000} = 16^{\rm h}03^{\rm m}25^{\rm s}$, 
               $\delta_{2000} = 66\degr 37\arcmin 31\arcsec$) 
and "b" ($\alpha_{2000} = 16^{\rm h}02^{\rm m}54^{\rm s}$, 
         $\delta_{2000} = 66\degr 41\arcmin 34\arcsec$)
as denoted by Montagni et al. (1996) were used as comparison 
stars for 
our CCD measurements. We measured their $U,~B,~V$ magnitudes 
with respect to our photoelectric standard BD+67$\degr$925 
(a: $V$ = 10.456$\pm 0.005$, 
    $B$ = 11.007$\pm 0.008$, 
    $U$ = 11.059$\pm 0.012$; 
 b: $V$ = 11.112$\pm 0.007$, 
    $B$ = 11.858$\pm 0.011$, 
    $U$ = 12.057$\pm 0.015$). 
These magnitudes agree within uncertainties with those measured 
by Henden \& Munari (2006). 
The LCs show flares, maxima of which repeat with a period 
of approximately 1 year. During these events the colour 
index $U - B < 0$, whereas during quiescence we observed 
$U - B \ge 0$ (Fig.~8). This suggests a significant increase 
of the nebular component of radiation during active phases. 

In 2006 June AG\,Dra began a massive outburst that is similar 
in profile, but stronger in brightness, to that from 1980-82. 
It reached the brightness maximum during 2006 September
($U \sim $8.0) and afterwards was declining gradually to
$U \sim 9$\,mag in spring of 2007. With the analogy to 
the 1980-82 active phase, the second eruption could be 
expected during the summer of 2007. 
%
%
\subsection{Draco\,C--1}

$B,~V,~R_{\rm C},~I_{\rm C}$ magnitudes of Draco\,C--1 
are in Table~12. 
We used the standard stars from Henden \& Munari (2000). There are 12 
calibrated stars in the field of Draco\,C$-$1 (12$\times$12 arcmin). 
The stars No.~4,~7,~9,~11 and 12, which were within the field 
of our 2-m telescope, were selected to calibrate the Draco C-1 
measurements. 
%

\subsection{RW\,Hya}

The \ubv measurements of RW\,Hya 
were carried out at SAAO between 2004 April 22 and May 02. 
Magnitudes are summarized in Table~13 and shown in Fig.~9 
together with those published previously by S+04. 
To compare the available data, which were obtained within 
a large time period (from 1990 to 2004), we plotted them against 
the orbital phase. We used the ephemeris for the inferior 
conjunction of the giant given by the solution of spectroscopic 
orbit as published by Schild et al. (1996). 
%

\subsection{SY\,Mus} 

The \ubv measurements of SY\,Mus 
are listed in Table~14. 
Observations were carried out at the SAAO on 2004 April. 
The star HD\,100264 (SAO\,251442; $V$ = 8.679, $B-V$ = 0.149, 
$U-B$ = -- 0.153) was used as a comparison star. 
%

\subsection{AR\,Pav}

The \ubv measurements of AR\,Pav 
are listed in Table~15. Stars 
  CD-66\,2195 (GSC 09080-01017: $V$ = 9.997, $B-V$ = 0.254, 
                                 $U-B$ = 0.164) 
and
  HD\,269743 (CPD-66\,331: $V$ = 10.514, $B-V$ = 0.781, 
                           $U-B$ = 0.378)
were used as standard stars. 
Figure~10 shows the recent evolution in the visual LC (from 1998.4 
to December 2006), which corresponds to a low stage of the AR\,Pav 
activity. Compared are also our \ubv measurements. 
The very good agreement between the visual estimates and 
photoelectric $V$ magnitudes suggests that variations 
in the visual LC with $\Delta m \ga 0.2$ reflect real changes. 
%
Below we point some interesting features in the LC: 

  (i) 
A 100$\div$150-day periodic variation developed between epochs 
69 and 70. This type of variability occurred sometimes during 
low levels of the activity. Skopal et al. (2000) ascribed 
it to pulsations of the red giant in the system. 

  (ii) 
A gradual decrease in the AR\,Pav activity is suggested by 
a slow decline in the $U$-band brightness. In 1999.6 we 
observed $U-V < 0$, while in 2004.3 $U-V > 0$, which reflects 
a decline of the nebular emission and thus also a decrease in 
the flux of the ionizing photons resulting probably from 
a temperature decrease of the hot source. 

  (iii) 
Positions of the recent two minima, 
  Min(69) = JD 2\,452966\,$\pm$\,2 
and
  Min(70) = JD~2\,453573.3\,$\pm$\,0.7, differ by 
607.3\,$\pm$\,2.1 days that is by 2.8 days larger then 
the orbital period ($P_{\rm orb}$ = 604.5 days). 
We ascribe this difference to a strongly variable size, 
geometry and radiation of the eclipsing object 
as already noted by Bruch et al. (1994). 

  (iv) 
Mid-points of all the minima from the low state of the activity 
(E = 66 to 70) precede those predicted by the linear ephemeris 
determined by 4$\div$65 epochs, which confirms a continuous 
decrease of the orbital period found by Skopal et al. (2000). 
%

\subsection{AG\,Peg}

Photometric observations of AG\,Peg 
are summarized in Table~16 and depicted in Fig.~11. 
The LCs show variations in the profile from cycle to cycle. 
Most pronounced are different levels of maxima/minima 
and their shaping. For example, the minimum around 
JD\,2\,452\,900 was flat for about 0.26 of the orbital 
period. 
Such behaviour suggests that the symbiotic nebula is variable
in both the shape and the emissivity. The former is given
by a different projection of the optically thick part of the
nebula into the line of sight (Skopal 2001), while the latter
can reflect variation in the flux of ionizing photons.

The variation in the $V$-band LC is probably in part 
cau\-sed by the giant's semiregular variability, because 
the light contribution from the giant dominates the SED from 
$V$  (cf. Fig.~20 of Skopal 2005). 
%

\subsection{AX\,Per}

The recent measurements of AX\,Per 
are introduced in Tables~17 and 18 and showed in Fig.~12. 
Our CCD frames from the pavilion G1 distinguished two optical 
components of the star BD+53$\degr$340 
($\alpha_{2000} = 01^{\rm h}36^{\rm m}37.98^{\rm s},~
  \delta_{2000} = 54\degr 14\arcmin 41.8\arcsec$) 
in \ubv\ filters. 
The brighter one (denoted as $a$) was used as a comparison
star for our CCD measurements in Table~18 
($a$: $V$ = 9.56, $B-V$ = 1.45 and 
 $b$: $V$ = 12.44, $B-V$ = 0.53). 
The wave-like profile of the LC as a function of the orbital 
phase reflects a quiescent phase. However, the profile is not 
simple sinusoidal. Variations during different orbital cycles 
are evident. 
%
%
%
%
\acknowledgements
The authors thank to Mr. Pavel Schalling and Mr. Kamil Kuziel
for taking photometric observations at the Skalnat\'{e} Pleso
observatory. Anonymous referee is thanked for useful comments 
that have improved the paper. 
This research was supported by a grant of the Slovak Academy 
of Sciences No. 2/7010/27 and by the Grant Agency of 
the Czech Republic, GACR 205/06/0217. 
%
%
%

%
%
%
%
%
%
\begin{figure*}
\centering
\begin{center}
\resizebox{16cm}{!}{\includegraphics[angle=-90]{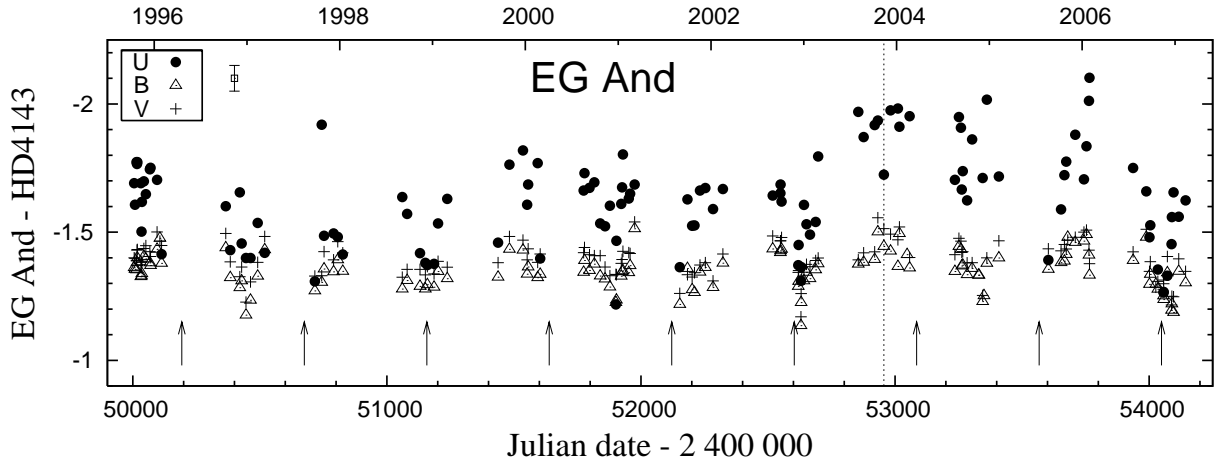}}
\caption{
 Differential \ubv LCs of EG\,And. Arrows mark positions of the 
 primary minima (ephemeris of Skopal 1997). New data (Table~1) 
 are plotted to the right of the vertical dotted line. 
 The error bar (top left) represents a maximum uncertainty 
 in $U$. 
}
\end{center}
\end{figure*}
%
%
%
\begin{figure*}
\centering
\begin{center}
\resizebox{16cm}{!}{\includegraphics[angle=-90]{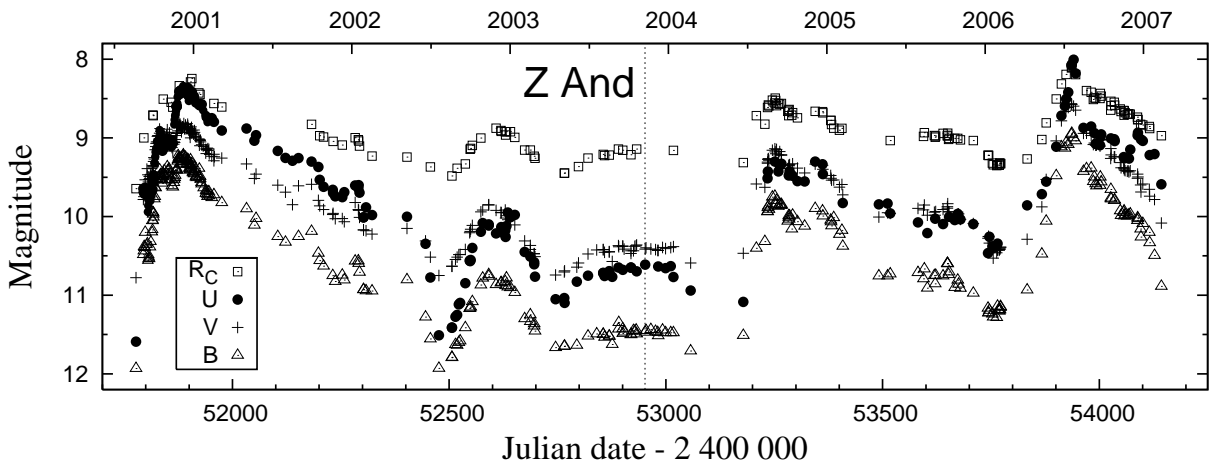}}
\caption{
 The \ubvr LCs of Z\,And covering its recent active
 phase from 2000. New data are from Tables~2 and 3. Compared 
 are data of Tomov et al. (2004) and Sokoloski et al. (2006) 
 around the 2000-01 maximum. 
}
\end{center}
\end{figure*}
%
%
%
\begin{figure*}
\centering
\begin{center}
\resizebox{16cm}{!}{\includegraphics[angle=-90]{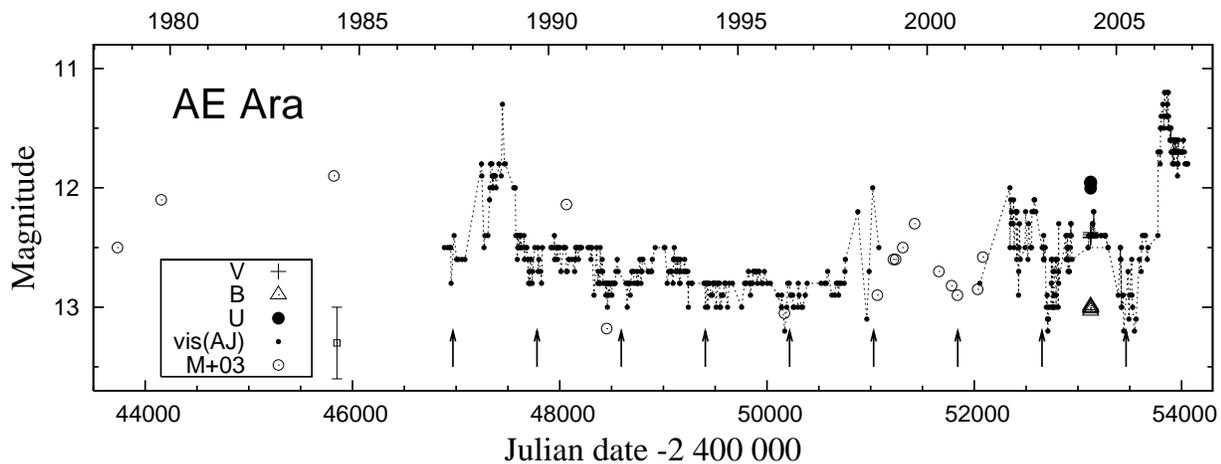}}
\caption{
  Our visual LC and \ubv magnitudes of AE\,Ara from Sect.~3.3 
  compiled with those published by Mikolajewska et al. (2003)
  (open circles). 
  Arrows denote positions of the primary minima according 
  to their ephemeris. New outburst began in 2006 February. 
The error bar represents a maximum uncertainty of faintest 
  visual estimates. 
}
\end{center}
\end{figure*}
%
%
%
\begin{figure*}
\centering
\begin{center}
\resizebox{16cm}{!}{\includegraphics[angle=-90]{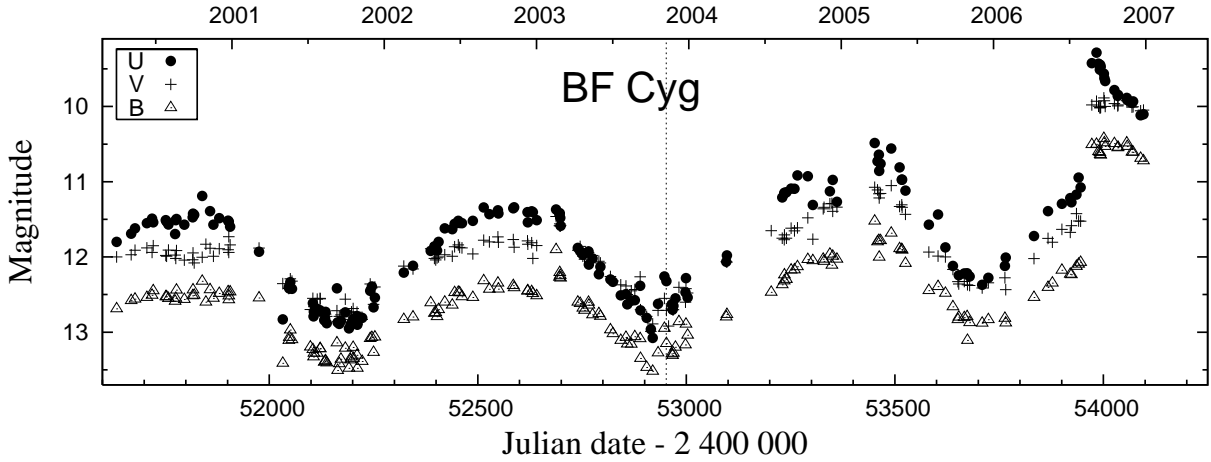}}
\caption{
The \ubv\ LCs of BF\,Cyg revealed a new active stage from 2006 August. 
Compared are data from Yudin et al. (2005) to 2004.6. 
}
\end{center}
\end{figure*}
%
%
%
\begin{figure*}
\centering
\begin{center}
\resizebox{16cm}{!}{\includegraphics[angle=-90]{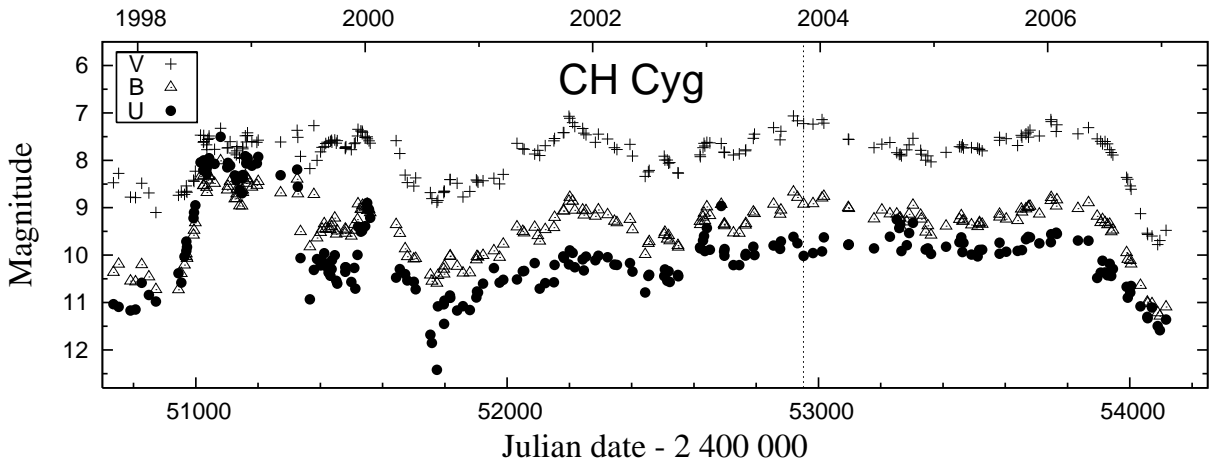}}
\caption{
 The \ubv LCs of CH\,Cyg. 
}
\end{center}
\end{figure*}
%
%
%
\begin{figure*}
\centering
\begin{center}
\resizebox{16cm}{!}{\includegraphics[angle=-90]{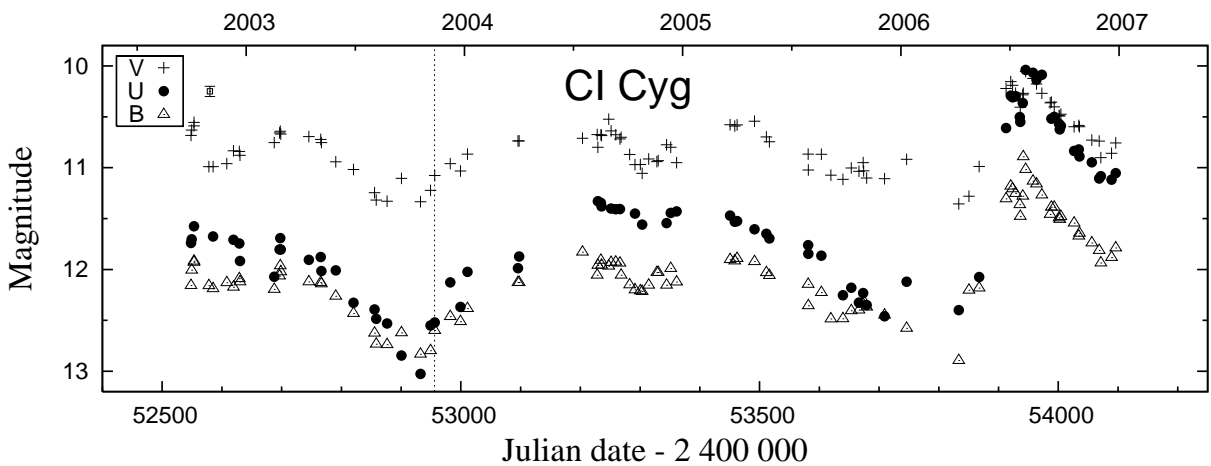}}
\caption{
 The \ubv LCs of CI\,Cyg revealed a new active stage that began 
 in 2006 June. Maximum uncertainty of individual points 
 is nearly within their size (compare the error bar at 
 the top-left corner). 
}
\end{center}
\end{figure*}
%
%
%
\begin{figure*}
\centering
\begin{center}
\resizebox{16cm}{!}{\includegraphics[angle=-90]{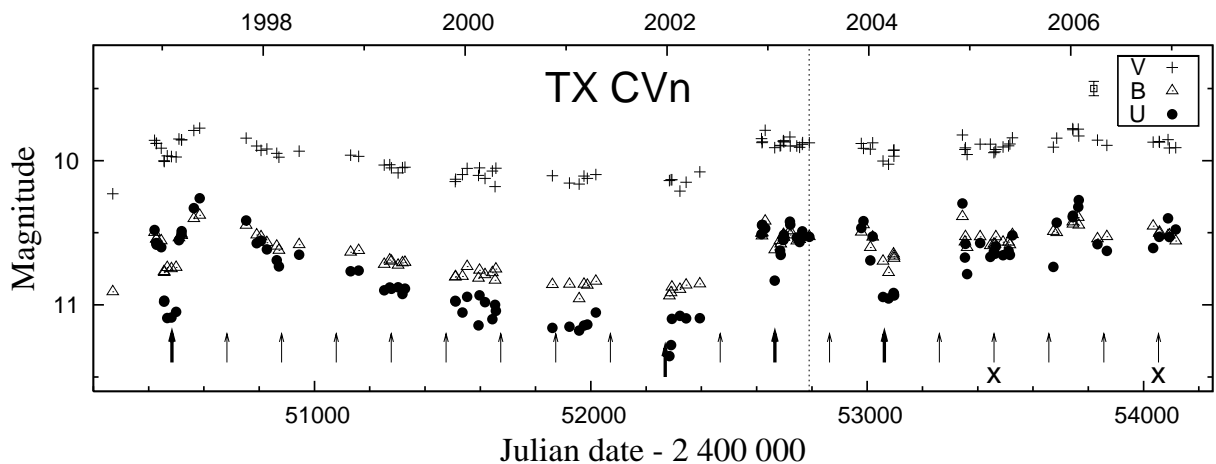}}
\caption{
 The \ubv LCs of TX\,CVn. Arrows denote positions of the inferior 
 conjunction of the giant (Kenyon \& Garcia 1989). Thick arrows 
 mark the appearance of minima, while $"\times"$ mark their 
 disappearance during the brighter stages (Sect.~3.8). 
 }
\end{center}
\end{figure*}
%
%
%
\begin{figure*}
\centering
\begin{center}
\resizebox{16cm}{!}{\includegraphics[angle=-90]{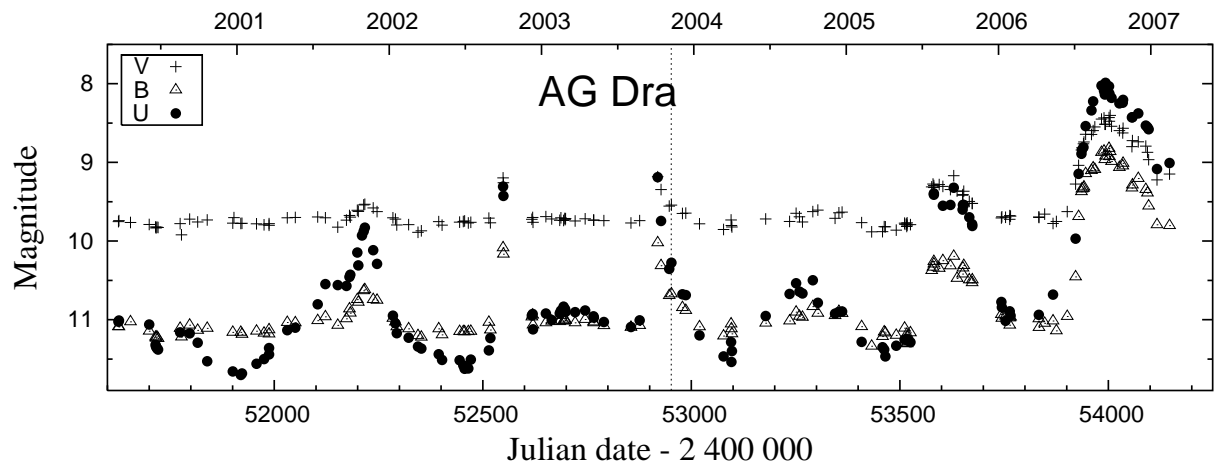}}
\caption{
 The \ubv LCs of AG\,Dra. New massive outburst started in 2006 June. 
 The data were complemented with those of Leedj\"arv et al. (2004). 
}
\end{center}
\end{figure*}
%
%
%
\begin{figure*}
\centering
\begin{center}
\resizebox{16cm}{!}{\includegraphics[angle=-90]{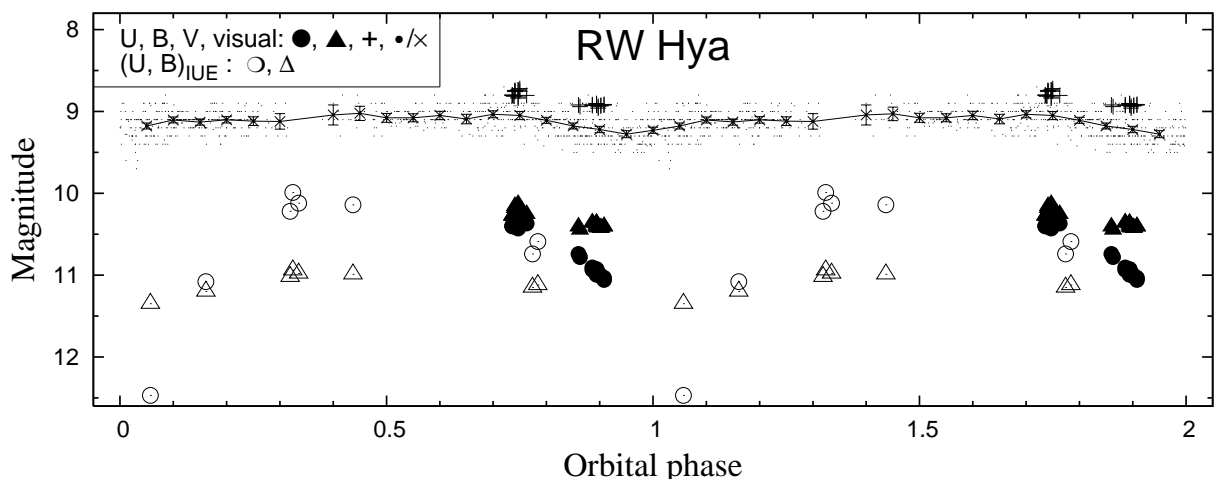}}
\caption{
  Our visual LC and \ubv measurements of RW\,Hya from Table~ 13 
  compiled with those published by S+04 and Munari et al. (1992). 
}
\end{center}
\end{figure*}
%
%
%
%
%
%
\begin{figure*}
\centering
\begin{center}
\resizebox{16cm}{!}{\includegraphics[angle=-90]{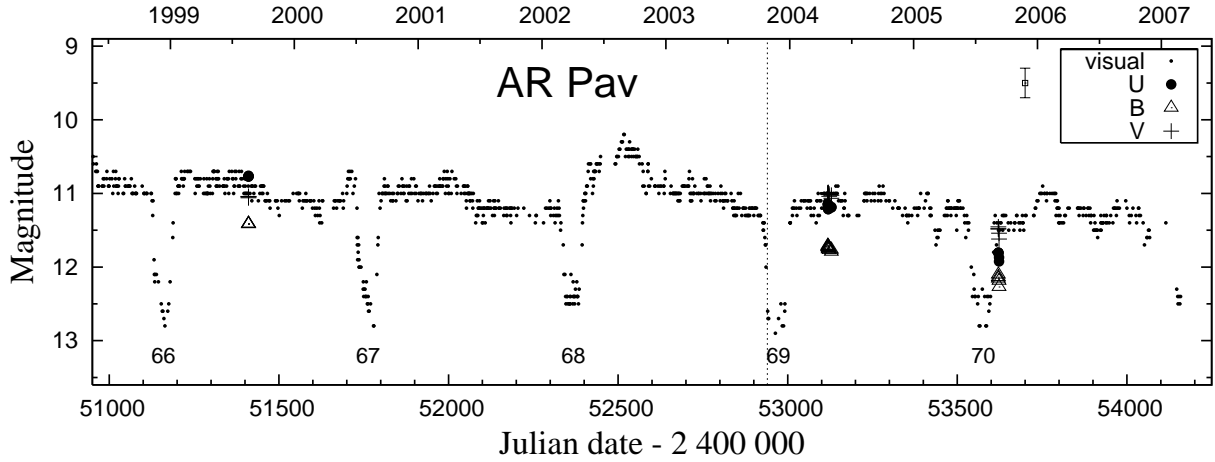}}
\caption{
  Our visual LC covering a low stage of activity from 1998.4. 
  New \ubv points are from Table~15. Numbers 66 -- 70 denote 
  the mid-points of eclipses predicted by their linear ephemeris 
  derived by Skopal et al. (2000). The error bar represents 
  a maximum uncertainty of visual estimates. 
}
\end{center}
\end{figure*}
%
%
%
\begin{figure*}
\centering
\begin{center}
\resizebox{16cm}{!}{\includegraphics[angle=-90]{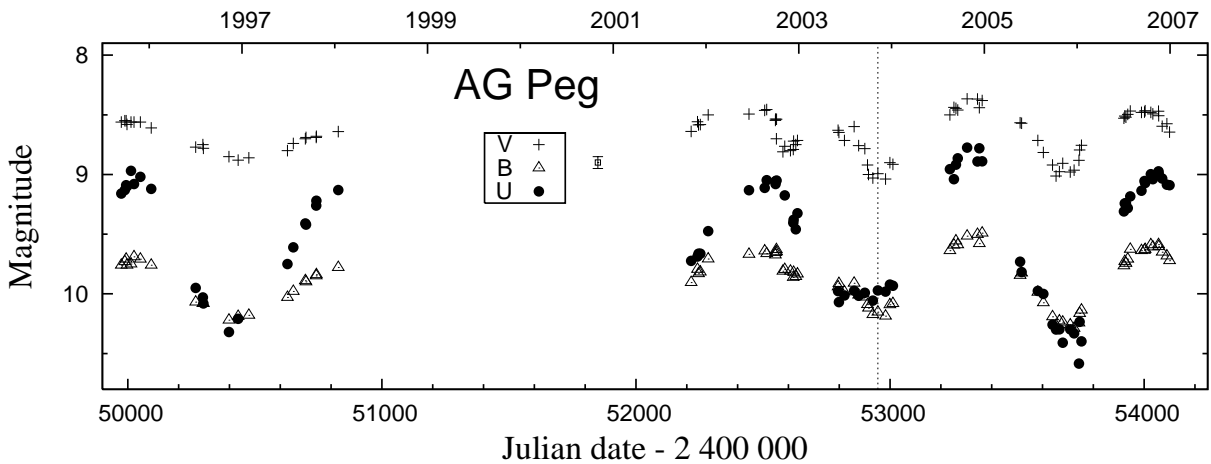}}
\caption{
  \ubv LCs of AG\,Peg. Data to 1998 are from Tomov \& Tomova (1998). 
}
\end{center}
\end{figure*}
%
%
\begin{figure*}
\centering
\begin{center}
\resizebox{16cm}{!}{\includegraphics[angle=-90]{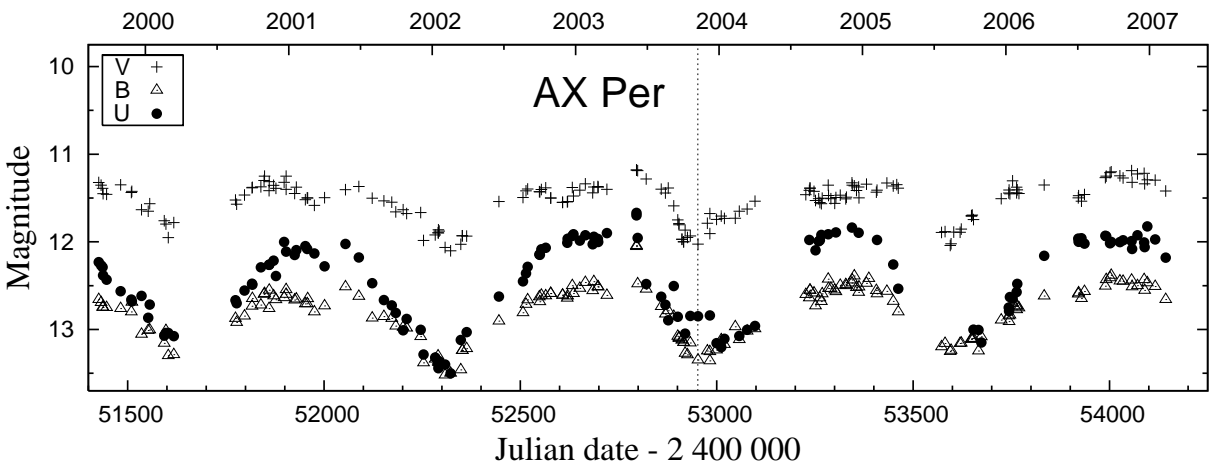}}
\caption{
  \ubv LCs of AX\,Per. New data are in Tables~17 and 18.
}
\end{center}
\end{figure*}
\clearpage
%
%
%
\begin{table*}
\begin{center}
\caption{$U,~B,~V,~R_{\rm C}$ observations of EG\,And.} 
\begin{tabular}{lccccccc}
\hline
\hline
   Date   & JD~24...   &$\Delta U$& $B$   &   $V$ &$\Delta R_{\rm C}$ 
                                                          & Obs \\
\hline
 Nov 12, 03 & 52956.229 & -1.724 &  8.718 &  7.049 & --     & G2 \\
 Dec  8, 03 & 52982.309 & -1.975 &  8.736 &  7.071 & --     & G2 \\
 Jan  6, 04 & 53011.312 & -1.982 &  8.796 &  7.093 & --     & G2 \\
 Jan 12, 04 & 53017.286 & -1.911 &  8.668 &  7.043 & -1.636 &SP  \\
 Feb 11, 04 & 53047.194 &  --    &  8.75  &  7.17  & --     & R \\
 Feb 21, 04 & 53057.256 & -1.952 &  8.802 &  7.162 & --     & G2 \\
 Aug 18, 04 & 53235.505 & -1.704 &  8.816 &  7.151 & --     & G2 \\
\hline
\hline
\end{tabular}
\end{center}
\normalsize
\end{table*}
%
%
\begin{table*}
\begin{center}
\caption{$U,~B,~V,~R_{\rm C}$ observations of Z\,And.}
\begin{tabular}{lcccccc}
\hline
\hline
Date & JD~24... & $U$ &  $B$   &  $V$ &$R_{\rm C}$& Obs\\
\hline
 Nov  8, 03 & 52952.222  & 10.612 & 11.472 & 10.414 & --     & G2  \\
 Dec  8, 03 & 52982.293  & 10.636 & 11.516 & 10.432 & --     & G2  \\
 Dec 25, 03 & 52999.275  & 10.656 & 11.478 & 10.408 & --     & G2  \\
 Jan  6, 04 & 53011.276  & 10.630 & 11.477 & 10.396 & --     & G2  \\
 Jan 12, 04 & 53017.246  & 10.769 & 11.479 & 10.383 &  9.160 &SP  \\
 Feb 21, 04 & 53057.229  & 10.940 & 11.710 & 10.590 & --     & G2  \\
 Jun 22, 04 & 53178.514  & 11.086 & 11.513 & 10.467 &  9.315 &SP  \\
\hline
\end{tabular}
\end{center}
\end{table*}
\addtocounter{table}{-1}
\begin{table*}
\begin{center}
\caption{Continued}
\begin{tabular}{lcccccc}
\hline
Date        & JD~24...   &    $U$ &   $B$  &   $V$  & $R_{\rm C}$&Obs\\
\hline
 Sep 13, 06 & 53991.528  &  9.076 &  9.584 &  9.094 &  8.494 &SP  \\
 Sep 14, 06 & 53993.361  &  9.072 &  9.611 &  9.092 & --     & G2  \\
 Sep 22, 06 & 54001.462  &  9.093 &  9.604 &  9.103 &  8.499 &SP  \\
 Sep 23, 06 & 54002.329  &  8.989 &  9.498 &  9.028 &  8.434 &SP  \\
 Sep 25, 06 & 54004.403  &  8.955 &  9.560 &  9.020 &  8.469 &SP  \\
 Oct 18, 06 & 54026.528  &  9.006 &  9.655 &  9.135 &  8.534 &SP  \\
 Oct 25, 06 & 54034.411  &  9.016 &  9.797 &  9.242 &  8.547 &SP  \\
 Oct 26, 06 & 54035.379  &  9.040 &  9.789 &  9.239 &  8.583 &SP  \\
 Nov 16, 06 & 54056.351  &  9.245 &  9.950 &  9.379 &  8.652 &SP  \\
 Nov 17, 06 & 54057.363  &  9.259 &  9.993 &  9.405 &  8.655 &SP  \\
 Nov 28, 06 & 54068.394  &  9.264 &  9.980 &  9.434 &  8.694 &SP  \\
 Dec  1, 06 & 54071.356  &  9.149 &  9.969 &  9.418 &  8.699 &SP  \\
 Dec 18, 06 & 54088.383  &  8.967 & 10.017 &  9.461 &  8.768 &SP  \\
 Dec 19, 06 & 54089.351  &  8.934 & 10.027 &  9.497 &  8.767 &SP  \\
 Dec 26, 06 & 54096.264  &  9.019 & 10.261 &  9.687 & --     & G2  \\
 Dec 30, 06 & 54100.270  &  9.039 & 10.090 &  9.585 &  8.837 &SP  \\
 Jan 15, 07 & 54116.231  &  9.219 & 10.334 &  9.741 &  8.880 &SP  \\
 Jan 26, 07 & 54127.217  &  9.205 & 10.495 &  9.783 & --     & G2  \\
 Feb 11, 07 & 54143.218  &  9.591 & 10.888 & 10.082 &  8.972 &SP  \\
\hline
\hline
\end{tabular}
\end{center}
\normalsize
\end{table*}
%
%
%
\begin{table*}
\begin{center}
\caption{$B,~V,~R_{\rm C},~I_{\rm C}$ CCD observations of Z\,And.} 
\begin{tabular}{lcccccc}
\hline
\hline
  Date     &  JD~24...   &  $B$   &  $V$   & $R_{\rm C}$ 
                                  & $\Delta I_{\rm C}$& Obs\\
\hline
 Feb 16, 03 & 52687.328  & 11.250 & 10.418 &   --   &   --   &G1  \\
 Jul 19, 03 & 52840.467  & 11.492 & 10.424 &   --   & -1.416 &G1  \\
 Sep  8, 03 & 52891.399  & 11.349 & 10.385 &   --   & -1.519 &G1  \\
 Sep 18, 03 & 52900.581  & 11.428 & 10.380 &   --   & -1.568 &G1  \\
 Sep 21, 03 & 52903.573  & 11.470 & 10.415 &   --   & -1.592 &G1  \\
 Oct 12, 03 & 52925.438  & 11.450 & 10.436 &   --   & -1.580 &G1  \\
 Oct 15, 03 & 52928.439  & 11.459 & 10.377 &   --   & -1.374 &G1  \\
\hline
\hline
\end{tabular}
\end{center}
\normalsize
\end{table*}
%
%
\begin{table*}
\begin{center}
\caption{$U,~B,~V,~R_{\rm C}$ observations of BF\,Cyg.} 
\begin{tabular}{lcccccc}
\hline
\hline
  Date   & JD~24...   &    $U$   &  $B$   & $V$ &$\Delta R_{\rm C}$ &Obs \\
\hline
 Nov 08, 03 & 52952.247 & 12.322 & 13.152 & 12.709 &  --    &G2  \\
 Dec  8, 03 & 52982.206 &   --   & 12.861 & 12.403 & --     & G2 \\
 Dec 25, 03 & 52999.185 & 12.281 & 12.894 & 12.422 & --     & G2 \\
 Mar 31, 04 & 53095.594 & 12.064 & 12.798 & 12.064 &  4.275 &SP  \\
 Apr  1, 04 & 53097.483 & 11.980 & 12.765 & 12.069 &  4.246 &SP  \\
 Jul 16, 04 & 53203.483 &  --    & 12.470 & 11.650 & --     & R  \\
 Aug 17, 04 & 53235.365 & 11.146 & 12.240 & 11.706 & --     & G2 \\
\hline
\end{tabular}
\end{center}
\end{table*}
\addtocounter{table}{-1}
\begin{table*}
\begin{center}
\caption{Continued}
\begin{tabular}{lcccccc}
\hline
Date        & JD~24...  &   $U$  &  $B$   &  $V$ &$\Delta R_{\rm C}$
                                                            & Obs\\
\hline
 Aug 24, 06 & 53972.321 &  9.425 & 10.508 &  9.980 & --     & G2  \\
 Sep  4, 06 & 53983.327 &  9.285 & 10.503 &  9.930 & --     & G2  \\
 Sep  9, 06 & 53988.309 &  9.431 & 10.601 &  9.998 &  2.627 &SP  \\
 Sep 12, 06 & 53991.378 &  9.511 & 10.612 & 10.018 &  2.660 &SP  \\
 Sep 13, 06 & 53992.279 &  9.447 & 10.639 & 10.016 & --     & G2  \\
 Sep 14, 06 & 53993.314 &  9.474 & 10.647 & 10.012 & --     & G2  \\
 Sep 22, 06 & 54001.254 &  9.565 & 10.425 &  9.887 &  2.640 &SP  \\
 Sep 23, 06 & 54002.256 &  9.625 & 10.483 &  9.923 &  2.650 &SP  \\
 Sep 25, 06 & 54004.287 &  9.661 & 10.531 & 10.000 &  2.717 &SP  \\
 Oct 17, 06 & 54026.279 &  9.781 & 10.495 &  9.962 &  2.740 &SP  \\
 Oct 25, 06 & 54034.212 &  9.860 & 10.550 &  9.992 &  2.800 &SP  \\
 Oct 26, 06 & 54035.259 &  9.847 & 10.543 &  9.979 &  2.734 &SP  \\
 Nov 16, 06 & 54056.186 &  9.887 & 10.486 &  9.923 &  2.708 &SP  \\
 Nov 17, 06 & 54057.257 &  9.915 & 10.532 &  9.919 &  2.720 &SP  \\
 Nov 28, 06 & 54068.272 &  9.953 & 10.607 & 10.009 &  2.825 &SP  \\
 Dec  1, 06 & 54071.184 &  9.932 & 10.609 & 10.003 &  2.769 &SP  \\
 Dec 19, 06 & 54089.207 & 10.116 & 10.690 & 10.057 &  2.769 &SP  \\
 Dec 26, 06 & 54096.181 & 10.104 & 10.723 & 10.049 & --     & G2  \\
\hline
\hline
\end{tabular}
\end{center}
\normalsize
\end{table*}
%
%
\begin{table*}
\begin{center}
\caption{$U,~B,~V,~R_{\rm C}$ observations of CH\,Cyg.}
\begin{tabular}{lcccccc}
\hline
\hline
Date        & JD~24...  & $U$    &   $B$  &   $V$  &$\Delta R_{\rm C}$ 
                                                            & Obs \\
\hline
 Nov  8, 03 & 52952.277 & 10.015 &  8.928 &  7.225 & --     & G2 \\
 Dec  8, 03 & 52982.269 &  9.956 &  8.917 &  7.237 & --     & G2 \\
 Jan  6, 04 & 53011.233 &  9.923 &  8.797 &  7.141 & --     & G2 \\
 Jan 12, 04 & 53017.200 &  9.627 &  8.763 &  7.216 & -2.667 &SP  \\
 Mar 31, 04 & 53095.511 &  9.783 &  8.999 &  7.554 & -2.545 &SP  \\
 Apr  1, 04 & 53097.398 &  9.773 &  9.020 &  7.561 & -2.411 &SP  \\
 Jun 21, 04 & 53178.464 &  9.857 &  9.237 &  7.735 & -2.211 &SP  \\
\hline
\end{tabular}
\end{center}
\end{table*}
\addtocounter{table}{-1}
\begin{table*}
\begin{center}
\caption{Continued}
\begin{tabular}{lcccccc}
\hline
Date        & JD~24...  & $U$    &  $B$   &  $V$   &$\Delta R_{\rm C}$ 
                                                            & Obs \\
\hline
 Jul 29, 06 & 53945.516 & 10.290 &  9.473 &  7.890 & -2.088 &SP  \\
 Sep 10, 06 & 53989.484 & 10.667 &  9.950 &  8.360 & -1.585 &SP  \\
 Sep 14, 06 & 53993.338 & 10.899 & 10.092 &  8.400 & --     & G2  \\
 Sep 22, 06 & 54001.346 & 10.790 & 10.116 &  8.542 & -1.406 &SP  \\
 Sep 25, 06 & 54004.358 & 10.650 & 10.190 &  8.612 & -1.348 &SP  \\
 Oct 25, 06 & 54034.327 & 11.080 & 10.641 &  9.125 & -0.857 &SP  \\
 Nov 16, 06 & 54056.316 & 11.301 & 10.992 &  9.539 & -0.410 &SP  \\
 Nov 17, 06 & 54057.220 & 11.331 & 11.022 &  9.566 & -0.393 &SP  \\
 Dec  1, 06 & 54071.312 & 11.105 & 11.029 &  9.597 & -0.411 &SP  \\
 Dec 19, 06 & 54089.281 & 11.496 & 11.219 &  9.777 & -0.322 &SP  \\
 Dec 26, 06 & 54096.238 & 11.583 & 11.289 &  9.694 & --     & G2  \\
 Jan 15, 07 & 54116.189 & 11.360 & 11.094 &  9.479 & -0.652 &SP  \\
\hline
\hline
\end{tabular}
\end{center}
\normalsize
\end{table*}
%
%
\begin{table*}
\begin{center}
\caption{$U,~B,~V,~R_{\rm C}$ observations of CI\,Cyg}
\begin{tabular}{lcccccc}
\hline
\hline
  Date      & JD~24...  & $U$    &  $B$   &  $V$   &$\Delta R_{\rm C}$
                                                            & Obs \\
\hline
 Nov 12, 03 & 52956.191 & 12.520 & 12.599 & 11.077 & --     & G2 \\
 Dec  8, 03 & 52982.246 & 12.127 & 12.464 & 10.960 & --     & G2 \\
 Dec 25, 03 & 52999.241 & 12.368 & 12.512 & 11.032 & --     & G2 \\
 Jan  6, 04 & 53011.185 & 12.024 & 12.384 & 10.866 & --     & G2 \\
 Mar 31, 04 & 53095.551 & 11.987 & 12.129 & 10.737 &  0.826 &SP  \\
 Apr  2, 04 & 53097.531 & 11.873 & 12.129 & 10.737 &  0.788 &SP  \\
 Jul 16, 04 & 53203.469 &  --    & 11.830 & 10.710 & --     & R  \\
\hline
\hline
\end{tabular}
\end{center}
\normalsize
\end{table*}
\clearpage
%
%
\begin{table*}
\begin{center}
\caption{CCD $B,~V,~R_{\rm C},~I_{\rm C}$ observations of CI\,Cyg.} 
\begin{tabular}{lcccccc}
\hline
\hline
   Date   & JD~24... &  $B$   &  $V$   &$\Delta R_{\rm C}$
                                              & $\Delta I_{\rm C}$& Obs\\
\hline
 Aug 10, 04&53228.325& 12.056 & 10.675 &0.578 & -0.937 &G1$^\dagger$\\
 Aug 29, 04&53247.376& 11.967 & 10.523 &0.748 & -0.486 &G1$^\ddagger$\\
 Sep 19, 04&53268.301& 12.056 & 10.701 &0.817 & -0.431 &G1$^\ddagger$\\
 Oct  3, 04&53282.421& 12.151 & 10.870 &0.982 & -0.396 &G1$^\ddagger$\\
 Oct 21, 04&53300.307& 12.210 & 10.968 &1.036 & -0.350 &G1$^\ddagger$\\
 Nov  4, 04&53314.320& 12.155 & 10.914 &1.027 & -0.379 &G1$^\ddagger$\\
 Sep  5, 05&53619.316& 12.486 & 11.074 &0.977 & -0.857 &G1$^\dagger$\\
 Apr 25, 06&53850.544& 12.204 & 11.280 &1.323 & -0.241 &G1$^\dagger$\\
 Jul 24, 06&53941.413& 10.893 & 10.269 &0.595 & -0.665 &G1$^\dagger$\\
 Sep  7, 06&53986.353& 11.459 & 10.363 &0.657 & -0.572 &G1$^\ddagger$\\
\hline
\hline
\end{tabular}
\end{center}
$^\dagger$CI\,Cyg - HD226107, $^\ddagger$CI\,Cyg - HD226041
\normalsize
\end{table*}
%
%
%
\begin{table*}
\begin{center}
\caption{CCD $B$ and $V$ observations of V1329\,Cyg from the Rozhen
         Observatory}
\begin{tabular}{lcccc}
\hline
\hline
  Date      &  JD~24... &  $B$  & $V$   & Obs \\
\hline
 Oct 02, 03 & 52915.371 & 13.69 & 13.00 & R  \\
 Jul 15, 04 & 53202.495 & 14.15 & 13.34 & R  \\
 Jul 16, 04 & 53203.499 & 14.15 & 13.33 & R  \\
 Nov 20, 04 & 53330.216 & 14.41 & 13.68 & R  \\
 Dec  8, 04 & 53348.199 & 14.49 & 13.78 & R  \\
 Dec  9, 04 & 53349.212 & 14.46 & 13.73 & R  \\
 Oct 30, 05 & 53674.246 & 14.36 & 13.46 & R  \\
\hline
\hline
\end{tabular}
\end{center}
\normalsize
\end{table*}
%
%
\begin{table*}
\begin{center}
\caption{$U,~B,~V,~R_{\rm C}$ observations of TX\,CVn}
\begin{tabular}{lcccccc}
\hline
\hline
   Date     & JD~24...  &  $U$   &  $B$   & $V$   & $\Delta R_{\rm C}$ 
                                                            & Obs \\
\hline
 May 30, 03 & 52790.374 & 10.529 & 10.527 &  9.877 &  0.106 &SP  \\
 Dec  5, 03 & 52978.678 & 10.467 & 10.496 &  9.881 &  0.050 &SP  \\
 Dec 13, 03 & 52986.655 & 10.420 & 10.445 &  9.915 &  0.147 &SP  \\
 Jan  6, 04 & 53011.451 & 10.692 & 10.603 &  9.921 & --     & G2 \\
 Jan 15, 04 & 53019.584 & 10.526 & 10.531 &  9.877 &  0.106 &SP  \\
 Feb 21, 04 & 53057.322 & 10.945 & 10.698 & 10.003 & --     & G2 \\
 Mar 12, 04 & 53077.415 & 10.956 & 10.777 & 10.025 &  0.288 &SP  \\
\hline
\hline
\end{tabular}
\end{center}
\normalsize
\end{table*}
%
%
\begin{table*}
\begin{center}
\caption{$U,~B,~V,~R_{\rm C}$ observations of AG\,Dra}
\begin{tabular}{lcccccc}
\hline
\hline
   Date    & JD~24...  & $U$ & $B$ & $V$ & $\Delta R_{\rm C}$
                                                           & Obs \\
\hline
 Nov  8, 03 & 52952.190 & 10.275 & 10.673 &  9.542 & --     &G2  \\
 Dec  5, 03 & 52978.629 & 10.677 & 10.846 &  9.648 & -0.781 &SP  \\
 Dec 13, 03 & 52986.601 & 10.688 & 10.883 &  9.644 & -0.759 &SP  \\
 Jan 15, 04 & 53019.539 & 11.200 & 11.090 &  9.780 & -0.640 &SP  \\
 Mar 12, 04 & 53077.481 & 11.466 & 11.207 &  9.852 & -0.598 &SP  \\
 Mar 30, 04 & 53095.467 & 11.285 & 11.055 &  9.731 & -0.745 &SP  \\
 Mar 31, 04 & 53096.332 & 11.537 & 11.117 &  9.801 & -0.557 &SP  \\
\hline
\end{tabular}
\end{center}
\end{table*}
\addtocounter{table}{-1}
\begin{table*}
\begin{center}
\caption{Continued}
\begin{tabular}{lcccccc}
\hline
Date        & JD~24...  & $U$    &  $B$   &  $V$   & 
                                               $\Delta R_{\rm C}$ & Obs \\
\hline
 Sep 12, 06 & 53991.451 &  8.137 &  8.965 &  8.514 & -1.595 &SP  \\
 Sep 14, 06 & 53993.258 &  7.988 &  8.921 &  8.515 & --     & G2  \\
 Sep 22, 06 & 54001.381 &  8.034 &  8.824 &  8.431 & -1.644 &SP  \\
 Sep 23, 06 & 54002.364 &  8.124 &  8.926 &  8.476 & -1.582 &SP  \\
 Sep 25, 06 & 54004.473 &  7.705 &  8.863 &  8.403 & -1.630 &SP  \\
 Sep 28, 06 & 54007.469 &  8.179 &  8.989 &  8.541 & -1.598 &SP  \\
 Oct 17, 06 & 54026.397 &  8.249 &  9.065 &  8.596 & -1.486 &SP  \\
 Oct 25, 06 & 54034.366 &  8.241 &  9.045 &  8.625 & -1.492 &SP  \\
 Oct 26, 06 & 54035.335 &  8.205 &  9.012 &  8.566 & -1.511 &SP  \\
 Nov 17, 06 & 54056.521 &  8.429 &  9.327 &  8.800 & -1.377 &SP  \\
 Nov 17, 06 & 54057.437 &  8.426 &  9.287 &  8.724 & -1.478 &SP  \\
 Dec  1, 06 & 54071.468 &  8.378 &  9.205 &  8.738 & -1.389 &SP  \\
 Dec 20, 06 & 54089.502 &  8.532 &  9.345 &  8.793 & -1.314 &SP  \\
 Dec 24, 06 & 54093.543 &  8.554 &  9.391 &  8.872 & -1.269 &SP  \\
 Dec 26, 06 & 54096.342 &  8.578 &  9.555 &  8.965 & --     & G2  \\
 Jan 15, 07 & 54116.352 &  9.084 &  9.794 &  9.223 & -0.990 &SP  \\
 Feb 14, 07 & 54146.466 &  9.008 &  9.804 &  9.148 & -1.071 &SP  \\
\hline
\hline
\end{tabular}
\end{center}
\normalsize
\end{table*}
%
%
\begin{table*}
\begin{center}
\caption{CCD $B,~V,~R_{\rm C},~I_{\rm C}$ observations of AG\,Dra. 
  Indices $a$ and $b$ denote the used comparison stars (Sect.~4.9)}
\begin{tabular}{lcccccc}
\hline
\hline
Date   &  JD~24...      &  $B$   &  $V$   &  
                       $\Delta R_{\rm C}$ & $\Delta I_{\rm C}$  &Obs \\
\hline
 Mar  3, 05 & 53432.504 & 11.338 &  9.883 & -0.945 & -1.454 &G1$^a$\\
 May 21, 05 & 53512.335 & 11.287 &  9.788 & -1.197 & -1.562 &G1$^a$\\
 May 25, 05 & 53516.338 & 11.306 &  9.782 & -1.811 & -2.128 &G1$^b$\\
 May 26, 05 & 53517.337 & 11.306 &  9.769 & -1.825 & -2.141 &G1$^b$\\
 Jul 20, 05 & 53572.422 &  --    &   --   & -2.184 & -2.508 &G1$^b$\\
 Jul 24, 05 & 53576.335 & 10.375 &  9.311 & -2.140 & -2.435 &G1$^b$\\
 Jul 27, 05 & 53579.367 & 10.333 &  9.276 & -2.214 & -2.465 &G1$^b$\\
 Aug 11, 05 & 53594.315 & 10.347 &  9.281 & -2.269 & -2.479 &G1$^b$\\
 Sep 23, 05 & 53637.303 & 10.475 &  9.365 & -2.223 & -2.420 &G1$^b$\\
 Jan 27, 06 & 53763.329 & 11.069 &  9.676 & -1.925 & -2.202 &G1$^b$\\
 Apr  8, 06 & 53833.566 & 11.099 &  9.701 &   --   & -2.131 &G1$^b$\\
 Apr 21, 06 & 53847.291 & 11.046 &  9.657 & -1.890 & -2.146 &G1$^b$\\
 May 19, 06 & 53875.445 & 11.141 &  9.750 & -1.854 & -2.121 &G1$^b$\\
 Jun 14, 06 & 53901.398 & 10.959 &  9.626 & -1.933 & -2.196 &G1$^b$\\
 Jul 24, 06 & 53941.340 &  9.314 &  8.741 & -2.589 & -2.717 &G1$^b$\\
 Aug 19, 06 & 53967.409 &  9.091 &  8.548 & -2.644 & -2.776 &G1$^b$\\
\hline
\hline
\end{tabular}
\end{center}
\normalsize
\end{table*}
%
%
%
\begin{table*}
\begin{center}
\caption{$B,~V,~R_{\rm C},~I_{\rm C}$ observations of Draco C-1}
\begin{tabular}{lcccccl}
\hline
\hline
 Date      &  JD~24... &  $B$   &  $V$  &$R_{\rm C}$& $I_{\rm C}$& Obs \\
\hline
Feb 17, 01 & 51957.626 &  --    & 17.12  &  --    &   --   & R 0.5-m \\
Mar 24, 01 & 51993.464 & 18.084 & 17.043 & 16.311 & 15.650 & R 2-m   \\
Mar 05, 02 & 52339.514 & 18.695 & 17.233 & 16.475 & 15.823 & R 2-m   \\
Feb 28, 03 & 52698.524 & 18.677 & 17.232 & 16.456 & 15,821 & R 2-m   \\
Mar 02, 03 & 52700.515 & 18.637 & 17.185 & 16.453 & 15.827 & R 2-m   \\
Mar 02, 03 & 52701.464 & 18.620 & 17.206 & 16.443 & 15.860 & R 2-m   \\
Apr 02, 03 & 52732.499 &  --    & 17.19  & 16.33  & 15.68  & R 0.5-m \\
Mar 12, 05 & 53442.474 & 18.619 & 17.133 & 16.341 & 15.679 & R 2-m   \\
\hline
\hline
\end{tabular}
\end{center}
\normalsize
\end{table*}
%
%
\begin{table*}
\begin{center}
\caption{$U,~B,~V$ observations of RW\,Hya}
\begin{tabular}{lccccc}
\hline
\hline
    Date & JD~24...     & $U$    &  $B$   &  $V$   & Obs \\
\hline
 Apr 22, 04 & 53118.377 & 10.401 & 10.275 &  8.802 & SAAO \\
 Apr 23, 04 & 53119.362 & 10.404 & 10.279 &  8.811 & SAAO \\
 Apr 26, 04 & 53122.406 & 10.427 & 10.298 &  8.829 & SAAO \\
 May 02, 04 & 53128.277 & 10.370 & 10.258 &  8.805 & SAAO \\
\hline
\hline
\end{tabular}
\end{center}
\normalsize
\end{table*}
%
%
\begin{table*}
\begin{center}
\caption{$U,~B,~V$ observations of SY\,Mus}
\begin{tabular}{lccccc}
\hline
\hline
   Date & JD~24...  & $U$ &   $B$   &   $V$  & Obs \\
\hline
 Apr 22, 04 & 53118.304 & 11.433 & 12.045 & 10.765 & SAAO\\
 Apr 22, 04 & 53118.305 & 11.434 & 12.058 & 10.762 & SAAO\\
 Apr 22, 04 & 53118.308 & 11.419 & 12.055 & 10.769 & SAAO\\
 Apr 22, 04 & 53118.309 & 11.425 & 12.045 & 10.763 & SAAO\\
 Apr 22, 04 & 53118.317 & 11.455 & 12.124 & 10.808 & SAAO\\
 Apr 22, 04 & 53118.318 & 11.439 & 12.135 & 10.816 & SAAO\\
 Apr 23, 04 & 53119.339 & 11.448 & 12.048 & 10.758 & SAAO\\
 Apr 23, 04 & 53119.339 & 11.429 & 12.065 & 10.751 & SAAO\\
 Apr 23, 04 & 53119.340 & 11.465 & 12.064 & 10.750 & SAAO\\
 Apr 23, 04 & 53119.341 & 11.464 & 12.055 & 10.760 & SAAO\\
 Apr 23, 04 & 53119.346 & 11.459 & 12.050 & 10.751 & SAAO\\
 Apr 23, 04 & 53119.347 & 11.453 & 12.057 & 10.750 & SAAO\\
\hline
\hline
\end{tabular}
\end{center}
\normalsize
\end{table*}
%
%
\begin{table*}
\begin{center}
\caption{$U,~B,~V$ observations of AR\,Pav}
\begin{tabular}{lccccc}
\hline
\hline
 Date &   JD~24... & $U$ & $B$ &  $V$  &  Obs \\
\hline
 Apr 23, 04 & 53118.549 & 11.144 & 11.710 & 11.004 & SAAO\\
 Apr 23, 04 & 53118.554 & 11.147 & 11.720 & 10.995 & SAAO\\
 Apr 23, 04 & 53118.557 & 11.151 & 11.729 & 10.996 & SAAO\\
 Apr 24, 04 & 53119.637 & 11.213 & 11.757 & 11.015 & SAAO\\
 Apr 24, 04 & 53119.637 & 11.214 & 11.742 & 11.005 & SAAO\\
 Apr 24, 04 & 53119.644 & 11.167 & 11.728 & 10.994 & SAAO\\
 Apr 24, 04 & 53119.645 & 11.165 & 11.720 & 10.988 & SAAO\\
 Apr 27, 04 & 53122.641 & 11.196 & 11.767 & 11.038 & SAAO\\
 Apr 27, 04 & 53122.647 & 11.178 & 11.742 & 11.010 & SAAO\\
 May  3, 04 & 53128.657 & 11.184 & 11.793 & 11.066 & SAAO\\
 May  3, 04 & 53128.662 & 11.192 & 11.765 & 11.025 & SAAO\\
 Sep  7, 05 & 53621.323 & 11.805 & 12.102 & 11.455 & SAAO\\
 Sep  7, 05 & 53621.329 & 11.808 & 12.157 & 11.484 & SAAO\\
 Sep  7, 05 & 53621.334 & 11.804 & 12.137 & 11.480 & SAAO\\
 Sep  9, 05 & 53623.299 & 11.870 & 12.187 & 11.540 & SAAO\\
 Sep  9, 05 & 53623.304 & 11.924 & 12.271 & 11.619 & SAAO\\
\hline
\hline
\end{tabular}
\end{center}
\normalsize
\end{table*}
%
%
%
\begin{table*}
\begin{center}
\caption{$U,~B,~V,~R_{\rm C}$ observations of AG\,Peg}
\begin{tabular}{lcccccc}
\hline
\hline
   Date     & JD~24...  &  $U$   &  $B$   &    $V$ &$\Delta R_{\rm C}$ 
                                                            & Obs \\
\hline
 Nov  8, 03 & 52952.301 &  9.972 & 10.154 &  8.993 & --     & G2 \\
 Dec  8, 03 & 52982.175 &  9.981 & 10.187 &  9.038 & --     & G2 \\
 Dec 25, 03 & 52999.215 &  9.922 & 10.092 &  8.901 & --     & G2 \\
 Jan  6, 04 & 53011.211 &  9.932 & 10.082 &  8.915 & --     & G2 \\
 Aug 17, 04 & 53235.449 &  8.956 &  9.638 &  8.501 & --     & G2 \\
 Sep  2, 04 & 53251.468 &  9.039 &  9.592 &  8.437 & -0.213 &SP  \\
 Sep 10, 04 & 53259.434 &  8.917 &  9.557 &  8.446 & -0.211 &SP  \\
\hline
\hline
\end{tabular}
\end{center}
\normalsize
\end{table*}
%
%
\begin{table*}
\begin{center}
\caption{$U,~B,~V,~R_{\rm C}$ observations of AX\,Per.} 
\begin{tabular}{lcccccc}
\hline
\hline
Date & JD~24...  & $U$ & $B$ & $V$ & $\Delta R_{\rm C}$ & Obs \\
\hline
 Nov  8, 03 & 52952.328 & 12.849 & 13.350 & 12.027 & --     & G2 \\
 Dec  8, 03 & 52982.326 & 12.841 & 13.359 & 11.906 & --     & G2 \\
 Dec 25, 03 & 52999.312 & 13.156 & 13.234 & 11.744 & --     & G2 \\
 Jan  6, 04 & 53011.259 & 13.203 & 13.119 & 11.705 & --     & G2 \\
 Jan 14, 04 & 53019.498 & 13.108 & 13.172 & 11.733 &  3.490 &SP  \\
 Feb 11, 04 & 53047.206 &  --    & 12.970 & 11.730 & --     & R  \\
 Feb 21, 04 & 53057.284 & 13.075 & 13.117 & 11.651 & --     & G2 \\
\hline 
\hline
\end{tabular}
\end{center}
\normalsize
\end{table*}
%
%
\begin{table*}
\begin{center}
\caption{$B$ and $V$ CCD observations of AX\,Per.}
\begin{tabular}{lcccc}
\hline
\hline
 Date       & JD~24...  &  $B$   &  $V$   &Obs \\
\hline
 Sep 21, 03 & 52903.617 & 13.103 & 11.791 & G1  \\
 Sep 21, 03 & 52903.640 & 13.093 & 11.805 & G1  \\
 Oct 12, 03 & 52925.472 & 13.290 & 11.949 & G1  \\
 Dec  2, 03 & 52976.292 & 13.246 & 11.788 & G1  \\
 Dec  8, 03 & 52982.376 & 13.253 & 11.677 & G1  \\
 Aug  9, 04 & 53226.589 & 12.601 & 11.462 & G1  \\
 Aug 20, 04 & 53237.579 & 12.555 & 11.386 & G1  \\
 Aug 20, 04 & 53237.595 & 12.552 & 11.390 & G1  \\
 Aug 30, 04 & 53247.509 & 12.584 & 11.423 & G1  \\
 Sep 19, 04 & 53268.393 & 12.594 & 11.502 & G1  \\
 Oct  6, 04 & 53284.539 & 12.534 & 11.472 & G1  \\
 Oct 11, 04 & 53290.479 & 12.534 & 11.507 & G1  \\
 Oct 21, 04 & 53300.404 & 12.550 & 11.561 & G1  \\
 Nov  4, 04 & 53314.375 & 12.498 & 11.467 & G1  \\
 Dec  5, 04 & 53345.177 & 12.437 & 11.353 & G1  \\
 Dec 22, 04 & 53362.166 & 12.517 & 11.499 & G1  \\
 Jan 10, 05 & 53381.349 & 12.475 & 11.341 & G1  \\
 Jan 16, 05 & 53387.283 & 12.419 &   --   & G1  \\
 Feb  4, 05 & 53406.259 & 12.556 & 11.434 & G1  \\
 Mar  3, 05 & 53433.235 & 12.566 & 11.328 & G1  \\
 Jul 20, 05 & 53572.448 & 13.198 & 11.892 & G1  \\
 Aug 12, 05 & 53594.510 & 13.252 & 12.041 & G1  \\
 Aug 14, 05 & 53596.509 & 13.238 & 12.022 & G1  \\
 Sep  6, 05 & 53619.529 & 13.160 & 11.890 & G1  \\
 Sep  8, 05 & 53622.444 & 13.156 & 11.855 & G1  \\
 Oct  5, 05 & 53648.580 & 13.107 & 11.701 & G1  \\
 Oct  5, 05 & 53648.616 & 13.114 & 11.707 & G1  \\
 Oct  7, 05 & 53650.508 & 13.106 & 11.695 & G1  \\
 Dec 19, 05 & 53724.303 & 12.894 & 11.508 & G1  \\
 Jan 27, 06 & 53763.302 & 12.735 & 11.377 & G1  \\
 Feb  2, 06 & 53769.264 & 12.754 & 11.449 & G1  \\
\hline
\hline
\end{tabular}
\end{center}
\normalsize
\end{table*}
\end{document}